\documentclass[aps,pre,twocolumn,showpacs,amsfonts,amssymb,floatfix]{revtex4-2}
\usepackage[T1]{fontenc}
\usepackage[utf8]{inputenc}
\usepackage{float}
\usepackage{color}
\usepackage{graphicx}
\usepackage{amssymb}
\usepackage{amsmath}
\usepackage{enumerate}
\usepackage{bbold}

\makeatletter

\date{April 11, 2023} 

\makeatother

\begin{document}

\title{Flocking in Binary Mixtures of Anti-aligning Self-propelled Particles}

\author{Rüdiger Kürsten}
\affiliation{Departament de Física de la Matèria Condensada, Universitat de Barcelona, Martí i Franquès 1, 08028 Barcelona, Spain}
\affiliation{Universitat de Barcelona Institute of Complex Systems (UBICS), 08028 Barcelona, Spain}
\affiliation{Institut für Physik, Universität Greifswald, Felix-Hausdorff-Str. 6, 17489 Greifswald, Germany}
\author{Jakob Mihatsch}
\affiliation{Institut für Physik, Universität Greifswald, Felix-Hausdorff-Str. 6, 17489 Greifswald, Germany}
\author{Thomas Ihle}
\affiliation{Institut für Physik, Universität Greifswald, Felix-Hausdorff-Str. 6, 17489 Greifswald, Germany}

\begin{abstract}
We consider two species of self-propelled point particles: A-particles and B-particles. The orientations between nearby particles are subject to pair interactions of different strength for A-A-, A-B-(=B-A-) and B-B-interactions, respectively. Even if all interactions involved are repelling, that is, if they locally favor anti-alignment between each pair of particles, we find global polar order of both A-particles and B-particles
We find qualitative agreement between agent-based simulations and mean field theory. Beyond mean field, we develop a Boltzmann-scattering theory based on one-sided molecular chaos that yields excellent quantitative agreement with simulations for dilute systems. 
For large systems, we find, depending on parameters, either micro-phase-separation or static patterns with either patches or stripes that carry different polarization orientations.
\end{abstract}
\maketitle

Active matter is characterized by an interplay of directed motion due to the consumption of free energy and dissipation, see e.g. \cite{SWWGR20, Ramaswamy17, MJRLPRS13, Ramaswamy10, SSBMV22, Chate20, ACJ22, BGHP20} for reviews.
Active systems arise naturally or engineered across scales from micro tubuli up to humans or robots.
Emergent phenomena in active systems are not subject to the laws of equilibrium statistical mechanics due to the constant driving.
Nevertheless, collective phenomena and non-equilibrium phase transitions can be observed.
One of the typical transitions of large collections of self-propelled particles is the so called flocking transition.
It was first studied in the famous Vicsek model of aligning self-propelled particles \cite{VCBCS95, TT98}, where the dynamics of the self propelled particles is formally given by a certain set of rules.
Similar models based on Langevin-equations have been studied later \cite{PDB08, FMMT12, CAP13, BCDDB13, MLDP18, CSP21}.

Due to the mesoscopic to macroscopic size of self-propelled particles they are not perfectly identical but certain properties of the dynamics differ across the population.
An important class of non-uniform collections of active particles are binary mixtures of two species of particles with different properties.
For aligning self-propelled particles they have been studied e.g. in \cite{Menzel12,CMWRN22, KK22}, where particles of the same species always align.
Between particles of different species, also anti-alignment has been considered.

In this letter we study binary mixtures of purely anti-aligning self-propelled particles by means of kinetic theory and agent-based simulations.
Surprisingly, we find that there is a flocking transition towards a globally polarised state despite the microscopic interactions favoring anti-alignment.
The flocking transition and flocking state can be qualitatively understood in mean field theory, that is also quantitatively correct in the limit of large velocities.
For smaller velocities and dilute systems we adopt a recently developed scattering theory \cite{IKL23A, IKL23B} based on one-sided molecular chaos in order to understand the flocking transition on a quantitative level.
The flocking state for large systems is non-homogeneous which we can understand by means of a mean field linear stability analysis of the (always unstable) homogeneous flocking state.
We find two types of instabilities that lead either to well known micro phase separated states \cite{SCT15} or to a novel type of patterns of locally polarized patches or stripes.

\textit{Model.}
We consider two species (A-particles and B-particles) of self-propelled point particles in two dimensions subject to periodic boundary conditions.
The particle positions are denoted by $\mathbf{r}_i=(x_i, y_i)$ and the direction of self-propulsion is described by the angle $\phi_i\in [-\pi, \pi]$. 
We denote the number of A-particles, the number of B-particles and the total number of particles by $N_A, N_B$ and $N$, respectively. 
The indexes of A- and B-particles are $\{1, \dots, N_A$ and $\{N_A+1, \dots, N\}$, respectively.
The time evolution is following the Langevin dynamics
\begin{align}
	\dot{x}_i&= v \cos \phi_i, \; \;
	\dot{y}_i= v \sin \phi_i,
	\notag
	\\
	\dot{\phi}_i&= \sum_{j\in\Omega_i}\Gamma_{ij} \cdot \sin(\phi_j-\phi_i) + \sigma \cdot \xi_i,
	\label{eq:langevin_dynamics}
\end{align}
where $\Omega_i=\{k: |\mathbf{r}_i-\mathbf{r}_k|<R\}$ denotes the set of indexes of neighbors of particle $i$, that is the indexes of all particles that are closer to particle $i$ than distance $R$. 
The angular dynamics is subject to independent Gaussian white noise terms $\xi_i$ with noise strength $\sigma\ge 0$.
The coupling matrix is symmetric and consists only of three different entries:
\begin{align}
	\Gamma_{ij}=\begin{cases} \Gamma_A &\text{if } i\in \{1, N_A\} \text{ and }j\in \{1, N_A\},\\
		\Gamma_{AB} &\text{if } i\in \{1, N_A\} \text{ and }j\in \{N_A+1, N\}\\
		&\text{or }  i\in \{N_A+1, N\} \text{ and }j\in \{1, N_A\},\\
	\Gamma_B  &\text{if } i\in \{N_A+1, N\} \text{ and }j\in \{N_A+1, N\}.\end{cases}
	\label{eq:coupling}
\end{align}
In this letter we consider only anti-aligning couplings $\Gamma_A, \Gamma_B, \Gamma_{AB} \le 0$.

\textit{Mean field theory.}
The Langevin dynamics \eqref{eq:langevin_dynamics} is equivalent to the Fokker-Planck equation
\begin{align}
	\partial_t P =& -v\sum_{i=1}^{N}cos \phi_i \partial_{x_i}P -v\sum_{i=1}^{N}sin \phi_i \partial_{y_i}P
	\notag
	\\
	&-\sum_{i=1}^{N}\partial_{\phi_i}\sum_{j\in \Omega_i} \Gamma_{ij} sin(\phi_j-\phi_i)P +\sum_{i=1}^{N}\frac{\sigma^2}{2}\partial_{\phi_i}^2 P,
	\label{eq:Fokker_Planck}
\end{align}
where $P$ denotes the probability density of the full system that depends on all spatial and angular coordinates $\{\mathbf{r}_i, \phi_i\}$ (arguments omitted for short notation).
For further simplifications we assume that the $N$-particle probability density factorizes into a product of $N_A$ identical and independent one-particle probabilities of A-particles and $N_B$ identical and independent one-particle probabilities of B-particles $P=P_A(\mathbf{r}_1, \phi_1) \cdot \dots \cdot P_A(\mathbf{r}_{N_A}, \phi_{N_A}) \cdot P_B(\mathbf{r}_{N_A+1}, \phi_{N_A+1})\cdot \dots \cdot P_B(\mathbf{r}_N, \phi_N)$.
Furthermore, we assume that the particles are distributed homogeneously, that is $P_{A/B}(\mathbf{r}, \phi)=\frac{1}{\mathcal{A}}p_{A/B}(\phi)$, where $\mathcal{A}$ is the area of the domain the particles move in.
For simplicity we consider only solutions that satisfy the symmetry $\phi \leftrightarrow -\phi$.\\
Plugging the above assumptions into Eq. \eqref{eq:Fokker_Planck}, integrating over all degrees of freedom but $\phi_1/\phi_{N_A+1}$ and renaming $\phi_1/\phi_{N_a+1}$ into $\phi$ we obtain
\begin{align}
	&\partial_t p_{A/B}(\phi)=\Gamma_{A/B} M_{A/B} \langle \cos \varphi\rangle_{A/B} \partial_{\phi} \sin \phi p_{A/B}(\phi)
	\notag
	\\
	&+\Gamma_{AB} M_{B/A} \langle \cos \varphi\rangle_{B/A} \partial_{\phi} \sin \phi p_{A/B}(\phi) + \frac{\sigma^2}{2}\partial_\phi^2 p_{A/B}(\phi),
	\label{eq:FPmeanfield}
\end{align}
where $M_{A/B}:=\pi R^2 N_{A/B}/\mathcal{A}$ is the expected number of neighboring A or B particles and for any function $f$, $\langle f\rangle_{A/B}:=\int_0^{2\pi} f(\alpha)p_{A/B}(\alpha) d\alpha$ is the expectation value of that function with respect to $p_{A/B}$.

\textit{Steady state solutions.}
Assuming that the expectation values $\langle \cos (\phi)\rangle_{A/B}$ are known a priori, Eq.~\eqref{eq:FPmeanfield} resembles overdamped equilibrium dynamics of variable $\phi$ in the potentials
\begin{align}
    V_{A/B}(\phi)=-(&\Gamma_{A/B} M_{A/B} \langle \cos \varphi\rangle_{A/B}
    \notag
    \\
    &+\Gamma_{AB} M_{B/A} \langle \cos \varphi\rangle_{B/A})\cos \phi
    \label{eq:potential}
\end{align}
with temperature $k_bT=\sigma^2/2$.
Thus the steady state is given by the Gibbs-Boltzmann-distribution.
For this particular potential the distribution is known as von Mises distribution, it reads
\begin{align}
    p_{A/B}(\phi)=\frac{1}{Z_{A/B}}exp\bigg[-\frac{2}{\sigma^2}V_{A/B}(\phi)\bigg]
    \label{eq:steadystatemf}
\end{align}
with normalization constant
\begin{align}
    Z_{A/B}=2\pi I_0\Big[\frac{2}{\sigma^2}(&\Gamma_{A/B} M_{A/B} \langle \cos \varphi\rangle_{A/B}
    \notag
    \\
    &+\Gamma_{AB} M_{B/A} \langle \cos \varphi\rangle_{B/A})\Big],
    \label{eq:normmf}
\end{align}
where $I_{\alpha}(x)$ is the modified Bessel function of first kind.

Note that the expectation values $\langle \cos \phi \rangle_{A/B}$ are not arbitrary but they have to satisfy the self-consistency condition
\begin{align}
    \langle \cos \phi \rangle_{A/B}= F_{A/B}(\langle \cos \phi \rangle_A, \langle \cos \phi \rangle_B),
    \label{eq:self-consistency-condition}
\end{align}
where the self-consistency map is given by
\begin{align}
    &F_{A/B}:=\int_0^{2\pi} d \phi \cos \phi p_{A/B}(\phi, \langle \cos \phi \rangle_A, \langle \cos \phi \rangle_B)
        \label{eq:self-consistency-map}
    \\
    &=\bigg\{I_1\Big[\frac{2}{\sigma^2}(\Gamma_{A/B} M_{A/B} \langle \cos \varphi\rangle_{A/B}
    \notag
    \\
    &+\Gamma_{AB} M_{B/A} \langle \cos \varphi\rangle_{B/A})\Big]\bigg\}
    \notag
    \\
    &\bigg/\bigg\{I_0\Big[\frac{2}{\sigma^2}(\Gamma_{A/B} M_{A/B} \langle \cos \varphi\rangle_{A/B}
    \notag
    \\
    &+\Gamma_{AB} M_{B/A} \langle \cos \varphi\rangle_{B/A})\Big]\bigg\}.
    \notag
    \label{eq:self-consistency-map}
\end{align}
We observe that the disordered state $\langle \cos \phi \rangle_{A/B}=0$ is always a solution of Eq.~\eqref{eq:self-consistency-condition}.
It changes its stability and bifurcates into nonzero solutions when the eigenvalue equation
\begin{align}
    \det \bigg[ \frac{\partial(F_A, F_B)}{\partial(\langle \cos \phi\rangle_A, \langle \cos \phi\rangle_B)}\bigg|_{\langle \cos \phi\rangle_{A/B}=0} -\lambda \mathbb{1}\bigg] =0
\end{align}
has a solution $\lambda^*= 1$.
Evaluating this condition we obtain the phase transition condition
\begin{align}
	\frac{1}{4}M_A M_B (\Gamma_A\Gamma_B-\Gamma_{AB}^2) 
 \notag
 \\
 - \frac{\sigma^2}{4}(\Gamma_A M_A+\Gamma_B M_B) + \frac{\sigma^4}{4}=0.
	\label{eq:ptc}
\end{align}
Note that this phase transition condition was obtained in \cite{Menzel12} (considering positive A-A and B-B couplings) from the dynamics of the zeroth and first angular Fourier modes only, without actually solving for the steady state distribution.
Remarkably, we find that there is a transition towards polar order even in the case of pure anti-aligning interactions, $\Gamma_{A/B}, \Gamma_{AB}<0$.
Expanding $p_{A/B}$ up to the second Fourier mode one can also show that the transition is continuous with the typical exponent $1/2$ within mean field.
Solving the self-consistency equation \eqref{eq:self-consistency-condition} in general analytically seems to be impossible.
However, it can easily be solved numerically by iterating the map \eqref{eq:self-consistency-map}.
Typically it converges after very few iterations.

\textit{Zero noise limit.}
Remarkably, for zero noise, the phase transition condition \eqref{eq:ptc} depends only on the ratio $\Gamma_{AB}^2/(\Gamma_A\Gamma_B)$.
That means it is independent on the densities $M_{A/B}$.
The zero noise limit allows to solve Eq.~\eqref{eq:self-consistency-condition} for the polar order parameters exactly, even in the ordered phase, leading to
\begin{align}
    |\langle \cos \phi \rangle_{A/B}|=\min\bigg(1, \frac{\Gamma_{AB} M_{B/A}}{\Gamma_{A/B}M_{A/B}}\bigg).
    \label{eq:polar_order_zero_noise}
\end{align}
If both polar order parameters are less than one, Eq.~\eqref{eq:polar_order_zero_noise} contradicts the condition of polar order $\Gamma_A\Gamma_B<\Gamma_{AB}^2$.
Thus, within the ordered state, there is always at least one species that is perfectly aligned while the other species is von Mises distributed with width given by \eqref{eq:polar_order_zero_noise}.
That implies in particular, that the transition is discontinuous in the singular case of zero noise within mean field theory.
\begin{figure}
    \centering
    \includegraphics[width=0.23\textwidth]{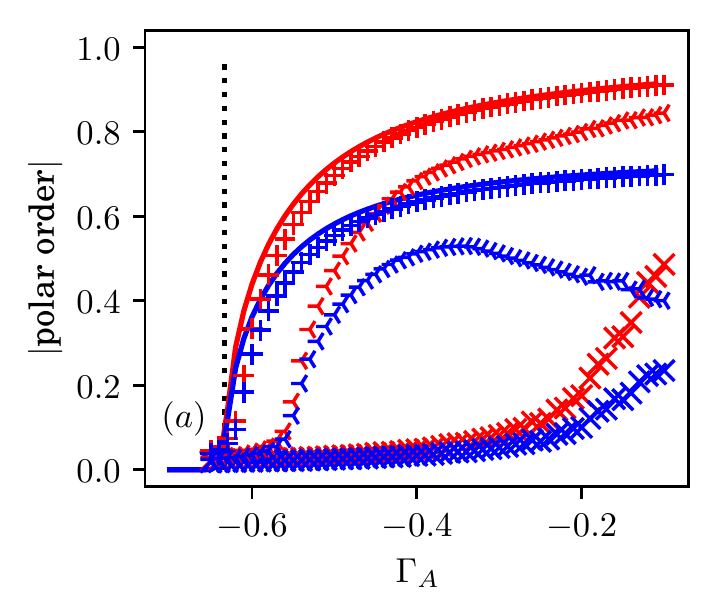}
    \includegraphics[width=0.23\textwidth]{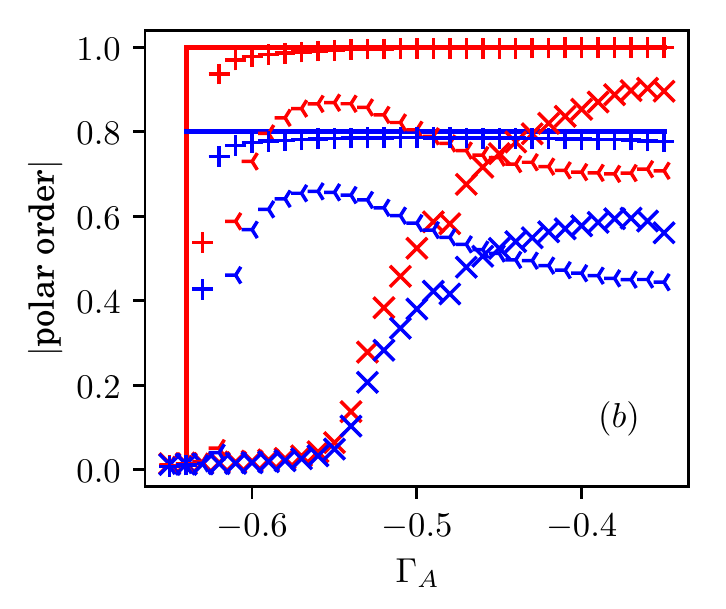}
    \caption{Absolute value of polar order parameter of A- (red) and B-particles (blue) for high densities and small system size ($N_A=N_B=2400$) with $(a)$ and without noise $(b)$. A- and B-particles are polarized in opposite directions. Agent-based simulations (symbols) are compared to mean field theory (solid line) [Eqs. \eqref{eq:steadystatemf}, \eqref{eq:polar_order_zero_noise}]. The dashed vertical black line in $(a)$ displays the mean field onset of flocking according to Eq. \eqref{eq:ptc}. Parameters are: $(a)$  $M_{A/B}=N_{A/B}\pi R^2/L^2 = 5$, $R=\sigma=1$, $\Gamma_B=\Gamma_{AB}=-1$. $(b)$ $M_{A/B}=N_{A/B}\pi R^2/L^2 = 20$, $R=1$, $\sigma=0$, $\Gamma_B=-1$, $\Gamma_{AB}=-0.8$. $v=1$ (cross-markers), $v=10$ (y-shaped markers), $v=100$ (plus-markers). Simulations have been
performed by AAPPP simulation package \cite{Kuersten22}. See \cite{supplemental} for numerical details.}
    \label{fig:polar_mf}
\end{figure}

We compare agent-based simulations to mean field theory in Fig. \ref{fig:polar_mf} with and without noise for large densities and different velocities.
The predicted presence of a flocking phase with oppositely polarized A- and B-particles is confirmed in agent-based simulations for all considered parameters.
For high velocity, the polar order parameter as well as the onset of flocking is coinciding well with mean field theory.
For smaller velocities, there is no quantitative agreement and the onset of flocking in simulations is shifted towards larger coupling compared to mean field predictions.

\textit{Beyond mean field: one-sided molecular chaos.}
Within mean field theory we assume that all particles are statistically independent at all times.
Due to the interactions this is obviously not strictly true.
However, the concept of molecular chaos gives an argument that correlations are vanishing fast due to collisions with many different particles.
Nevertheless, correlations can not be neglected on a quantitative level, cf. e.g. \cite{KSZI20, KI21}.

In order to improve upon mean field theory and incorporate the most relevant correlations in our theory we apply the concept of one-sided molecular chaos.
The idea is that two particles are assumed to be completely independent before they collide because they have most likely collided with many different particles before and lost almost all memory of a potential previous collision.
The collision between two particles however, lasts a finite period of time.
During this collision time, particles clearly build up correlations that can not be neglected on a quantitative level.
Following \cite{IKL23A, IKL23B} we take these collision correlations rigorously into account.
Here, we focus on dilute systems such that collisions between more than two particles at the same time can be neglected.
Furthermore, we assume that $v/R \gg \sigma^2$ 
such that the impact of noise on the collision is negligible.
In that way the collision dynamics is a two-body problem that is analytically manageable.
With  those ingredients we build a Boltzmann-like scattering theory in complete analogy to \cite{IKL23A, IKL23B} where the corresponding theory was derived for a single-species system.
As a result we obtain a Boltzmann-equation similar to \eqref{eq:FPmeanfield}.
Because it seems not analytically solvable, we Fourier transform the angular degrees of freedom according to
\begin{align}
    p_{A/B}(\phi)= \sum_{k} \hat{p}_{k}^{A/B}\exp(ik\phi),
\end{align}
yielding
\begin{align}
    &\partial_t \hat{p}^{A/B}_m=M_{A/B}\Gamma_{A/B} \Big[ \pi m (\hat{p}^{A/B}_{m-1}\hat{p}^{A/B}_{1} - \hat{p}^{A/B}_{m+1}\hat{p}^{A/B}_{-1})
    \notag
    \\
    &+\Gamma_{A/B}\frac{R}{\pi v}\sum_n\hat{p}^{A/B}_{m-n}\hat{p}^{A/B}_{n}g_{mn} \Big]
    \notag
    \\
    &+M_{B/A}\Gamma_{AB} \Big[ \pi m (\hat{p}^{A/B}_{m-1}\hat{p}^{B/A}_{1} - \hat{p}^{A/B}_{m+1}\hat{p}^{B/A}_{-1})
    \notag
    \\
    &+\Gamma_{AB}\frac{R}{\pi v}\sum_n\hat{p}^{A/B}_{m-n}\hat{p}^{B/A}_{n} g_{mn} \Big] -m^2\frac{\sigma^2}{2}\hat{p}^{A/B}_m,
    \label{eq:boltzmann_fourier}
\end{align}
where
\begin{align}
    g_{mn}=\frac{8}{3}m\bigg( \frac{\frac{3}{2}m-n}{n^2-\frac{9}{4}}+\frac{\frac{1}{2}m+n}{n^2-\frac{1}{4}}\bigg).
\end{align}
The terms that come with the factor $g_{mn}$ are corrections compared to mean field.
Neglecting modes of second and higher order we can analyze the linear stability of the disordered state yielding the phase transition condition
\begin{align}
    0=&\left(\Gamma_A -\frac{128}{45\pi^2} \Gamma_A^2 \frac{R}{v} -\frac{64}{9\pi^2} \Gamma_{AB}^2 \frac{M_B}{M_A} \frac{R}{v} -\frac{\sigma^2}{M_A}\right)
    \notag
    \\ 
    & \times\left(\Gamma_B -\frac{128}{45\pi^2} \Gamma_B^2 \frac{R}{v} -\frac{64}{9\pi^2} \Gamma_{AB}^2 \frac{M_A}{M_B} \frac{R}{v} -\frac{\sigma^2}{M_B}\right)
    \notag
    \\
    & -\left(\Gamma_{AB} +\frac{64}{15\pi^2} \Gamma_{AB}^2 \frac{R}{v}\right)^2 
    \label{eq:ptc2}
\end{align}
that is the analog to Eq.~\eqref{eq:ptc}.

\begin{figure}
    \centering
    \includegraphics[width=0.23\textwidth]{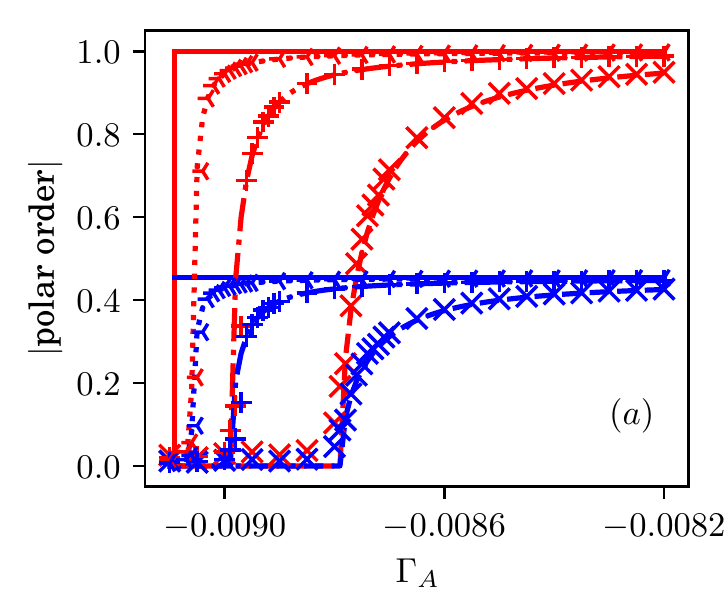}
    \includegraphics[width=0.23\textwidth]{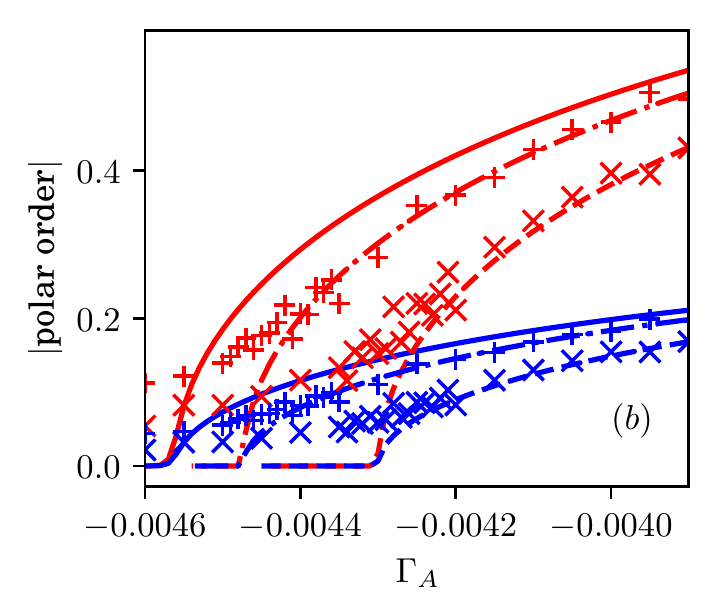}
    \caption{Polar order parameter of A- (red) and B-particles (blue). Agent-based simulations (symbols) are compared to mean field theory (solid line) [Eqs. \eqref{eq:steadystatemf}, \eqref{eq:polar_order_zero_noise}] and one sided molecular chaos based scattering theory [Eq. \eqref{eq:boltzmann_fourier}] for $v=1$ (dashed line), $v=3$ (dash-doted line) and $v=10$ (doted line).
    $(a)$ $\sigma=0$, $(b)$ $\sigma=0.01$. Parameters: $M_A=N_A\pi R^2/L^2=0.03$, $M_B=N_B\pi R^2/L^2=0.06$, $\Gamma_B=-0.011$, $\Gamma_{AB}=-0.01$, $v=1$ ($\times$), $v=3$ ($+$), $v=10$ (y-shaped marker). System size in simulations is given by $N_A=1600$.
    See \cite{supplemental} for numerical details.}
    \label{fig:bm_theory_sim}
\end{figure}

We integrate Eq. \eqref{eq:boltzmann_fourier} numerically taking modes up to $m=100$ into account.
In Fig. \ref{fig:bm_theory_sim} we compare the results of the scattering theory with agent-based simulations of small systems ($N_A=1600$, $N_B=3200$) for different velocities.
We find excellent quantitative agreement between theory and simulation regarding both, the polar order and the onset of flocking.
In particular, in contrast to mean field, the predictions based on one-sided molecular chaos reproduce the velocity dependence correctly.
Furthermore, not only the steady state is predicted correctly, but also the full dynamics of the orientational distribution, see \cite{supplemental}.
\begin{figure*}
    \centering
    \includegraphics[width=\textwidth]{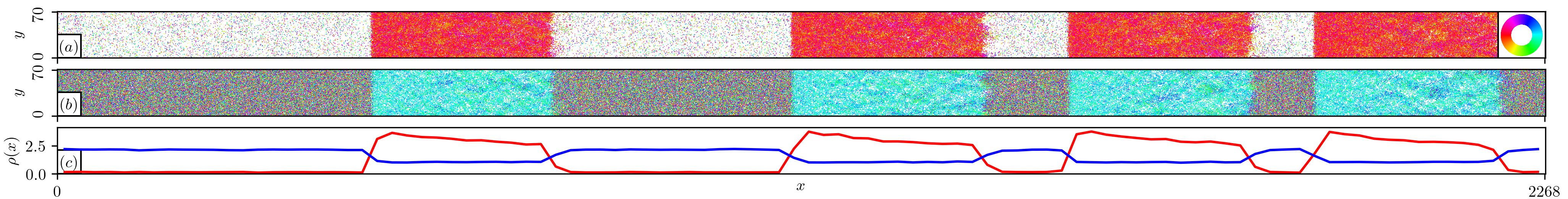}
    \caption{Snapshot of A-particles (a) and B-particles (b) in the micro-phase-separated state. Orientations are encoded in color according to the color wheel in (a).
    The local density of A-particles (red) and B-particles (blue) are displaced in (c).
    The system micro-phase-separates into polarly ordered regions with high density of A-particles and low density of B-particles, and a disordered regions with high density of B-particles and low density of A-particles.  See \cite{supplemental} for simulation details.}
    \label{fig:bands}
\end{figure*}

\begin{figure}
    \centering
    \includegraphics[width=0.23\textwidth]{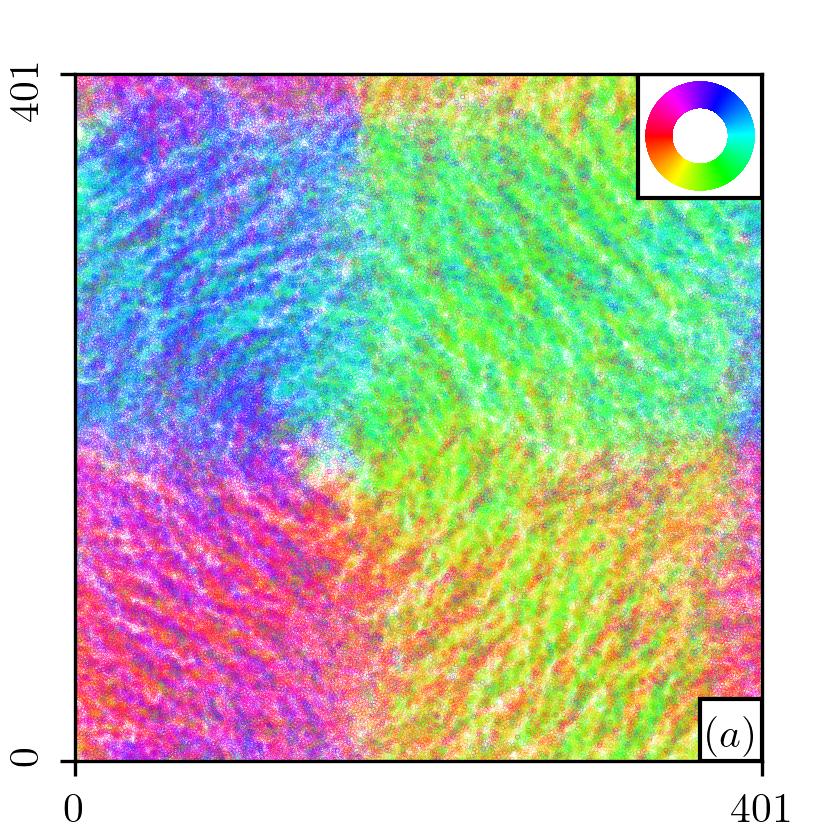}
    \includegraphics[width=0.23\textwidth]{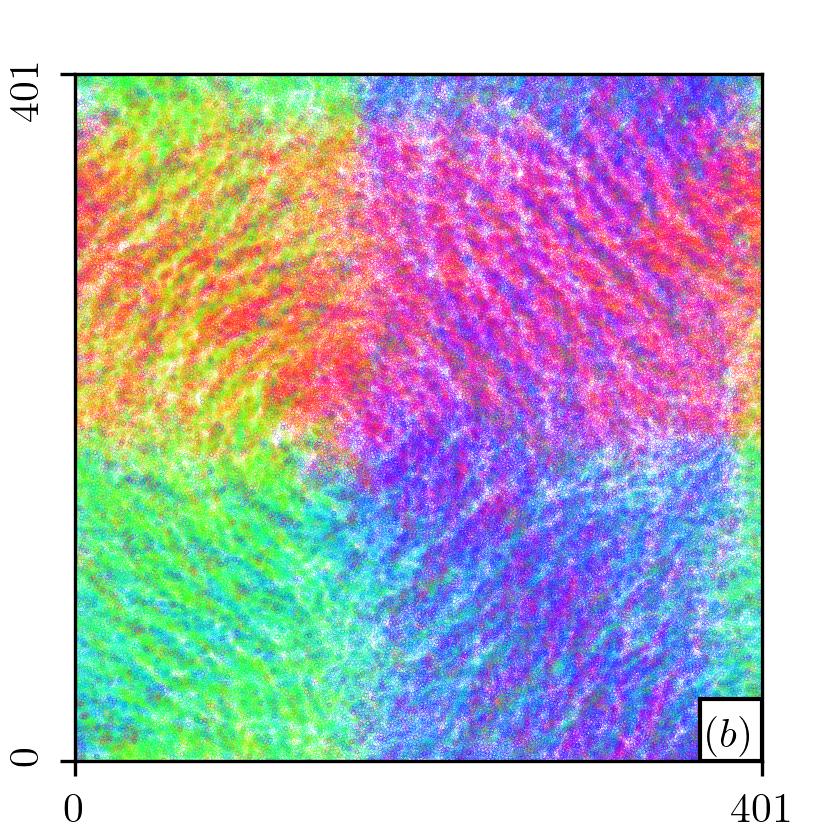}
    \includegraphics[width=0.23\textwidth]{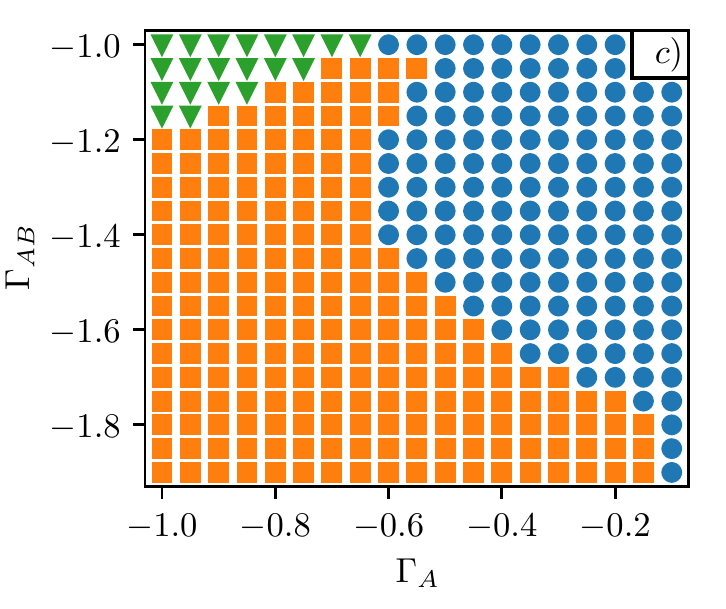}
    \includegraphics[width=0.23\textwidth]{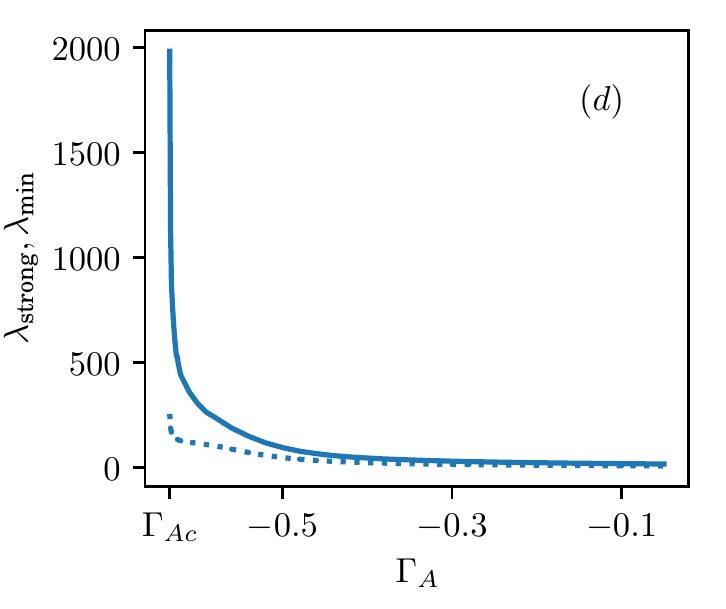}
    \caption{Snapshots of A-particles $(a)$ and B-particles $(b)$ in the phase of polarized patches. $(c)$ mean field phase diagram with disorder (green triangles), polarized moving bands (blue circles) and polarized patches (orange squares). $(d)$ wavelength of most unstable mode (solid line) and the smallest wavelength of all unstable modes (doted line) within the phase of polarized moving bands. Both wavelengths diverge when approaching the transition towards disorder. See \cite{supplemental} for simulation details.}
    \label{fig:patches}
\end{figure}

\textit{Spatially inhomogeneous states.} So far, we studied small systems and assumed a homogeneous spatial distribution of particles.
For aligning active particles it is known that spatial structures are formed within the flocking phase of large systems \cite{SCT15,KI21}.
Here, for anti-aligning active particles, we also find spatial structures within the flocking phase for large systems.
A mean field linear stability analysis reveals that the homogeneous disordered state is always stable and the homogeneous polarly ordered state is always unstable, see \cite{supplemental} for more details.
However, depending on parameters, there are two kinds of instabilities.
The first instability is purely longitudinal with respect to the polarization axis in positional space and it is symmetric with respect to reflections of orientations along the polarization axis.
Thus, the instability is not affecting the direction of the polarization.
This instability results in the formation of polarized bands, see Fig. \ref{fig:bands}.
The simulation domain micro-phase-separates into stripes of disorder and of high polar order, with A- and B-particles being polarized in opposite directions.
In this phase, particles partially demix: the density of A-particles is much higher in the polarized regions whereas the density of B-particles is much higher in the disordered regions.
The species with larger density within the polarized region dominates the dynamics of the pattern.
That means the polarized bands move into the direction of motion of the A-particles.
It should be mentioned that we observed long living meta stable dynamical patterns that eventually decay into the micro-phase-separated states, see \cite{supplemental} for details.

The second instability is purely transversal with respect to the polarization axis, however the angular modes of the unstable eigenvector are not symmetric with respect to reflections along the polarization axis.
Thus, the instability causes some bending of the polarization direction.
As a result of this instability we observe patterns of patches with different polarization orientation, see Fig. \ref{fig:patches} $(a-b)$.
In a few realizations, we also observed a pattern of stripes with different polarization orientations, see \cite{supplemental} for details.

In Fig. \ref{fig:patches} $(c)$ we show the mean field phase diagram with the three phases: disorder, micro-phase-separation and spatially inhomogeneous polarization patterns (patches or stripes).
It is worth mentioning that the wavelength of both instabilities diverges when approaching the disordered state.
We show this behavior exemplary for the transition from micro-phase-separation to disorder in Fig. \ref{fig:patches} $(d)$.
As a result, we observe homogeneous polarized states in simulations of finite size close to the onset of flocking.
We did not find a Toner-Tu phase for purely negative couplings, thus we suspect that this phase is only present in case that at least one of the couplings is positive.

In summary, we study a binary mixture (A- and B-particles) of anti-aligning self-propelled point particles in two dimensions.
Despite anti-aligning torques between each pair of particles we find a flocking state where A- and B-particles move in opposite directions.
Mean field theory correctly predicts the observed onset of flocking on a qualitative level and reasonable quantitative agreement is reached for very high velocities.
For small systems the flocking transition is continuous and the flocking states are homogeneous.
Only in the singular zero noise limit, a discontinuous transition is falsely predicted by mean field.
In reality the transition remains continuous even for zero noise due to correlation effects.
For dilute systems we develop a scattering theory based on one sided molecular chaos.
In that way we incorporate the major aspects of pair correlations during collisions.
The resulting theory agrees excellently with agent-based simulations on a quantitative level, not only regarding the steady state but it also covers the dynamics of the orientational distribution.
In particular, the correct dependence of the onset of flocking on the velocity is predicted and the transition of small systems is predicted to be continuous even in the noise free case.

For large systems we observe two types of patterns depending on parameters.
In the first case we observe micro-phase-separation between a disordered gas and polarized, moving bands that is known also for aligning active particles.
In the second case we observe a patterns of different polarization in different places either organized in quadratic patches or in stripes.
We understand the arising patterns by means of a linear stability analysis of the homogeneous flocking state within mean field theory and beyond hydrodynamics.
Considering all angular modes of the homogeneous flocking state and arbitrary many angular modes of the perturbation we find two types of instabilities of the homogeneous flocking state that are consistent with the observed patterns.

\begin{acknowledgments}
The authors thank Universitätsrechenzentrum Greifswald for supporting this work by providing computational resources.
R.K. acknowledges funding through a ’María Zambrano’ postdoctoral grant at University of Barcelona financed by the Spanish Ministerio
de Universidades and the European Union (Next Generation EU/PRTR).
\end{acknowledgments}

%

\begin{appendix}
\onecolumngrid
\newpage
\section{Scattering theory based on one-sided molecular chaos}

\begin{figure}[b]
    \centering
    \includegraphics[width=0.48\textwidth]{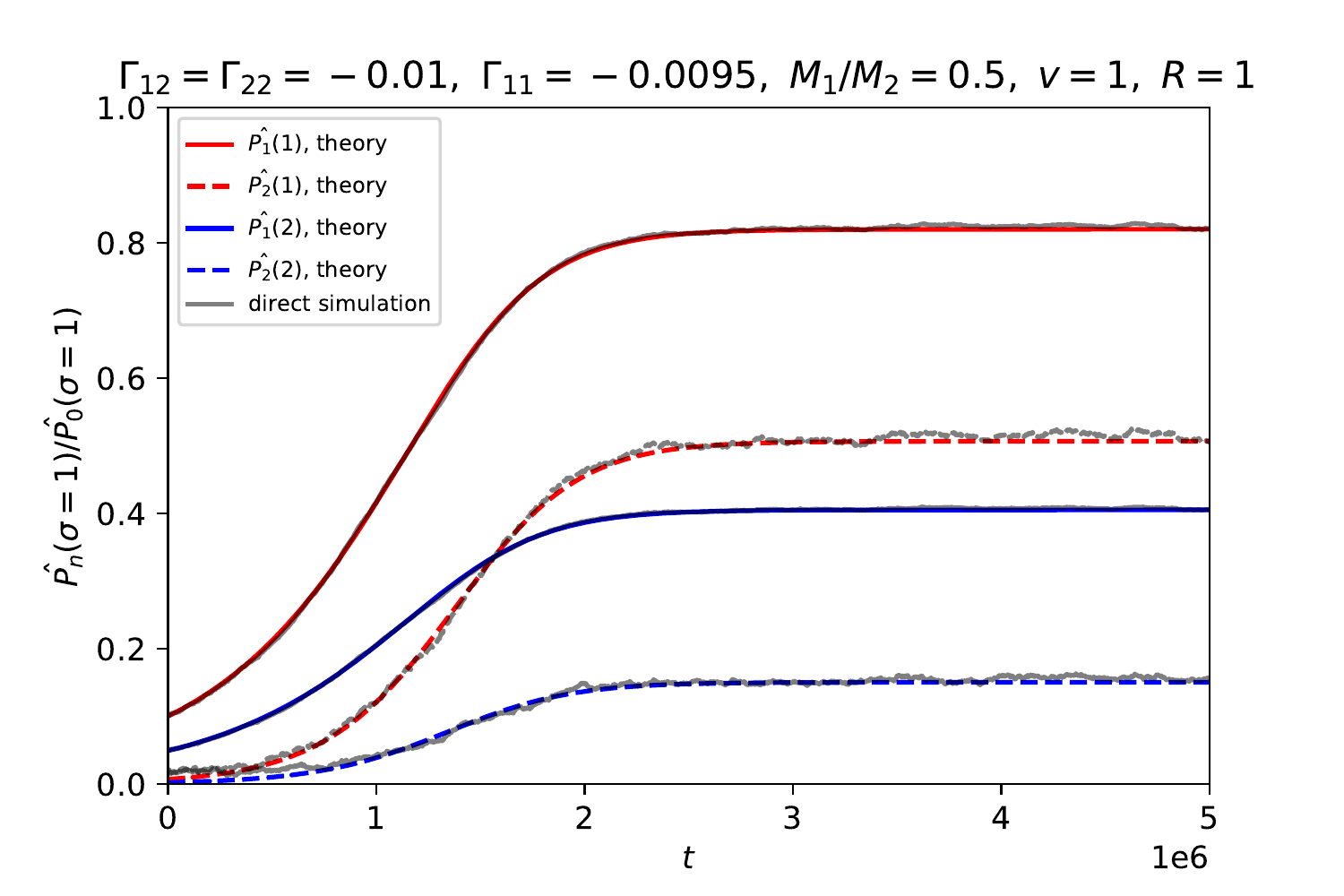}
    \caption{Comparison of the time evolution of the first two orientational Fourier modes between scattering theory and agent based simulations starting from initial conditions with nonzero polar order.}
    \label{fig:time_evolv_scatt}
\end{figure}

We have shown in Fig. \ref{fig:bm_theory_sim} that the steady state of agent-based simulations agrees excellent with the scattering theory presented in Eq. \eqref{eq:boltzmann_fourier} for dilute systems.
In fact, this quantitative agreement is not restricted to the steady state but Eq. \eqref{eq:boltzmann_fourier} describes the full dynamics correctly.
In Fig. \ref{fig:time_evolv_scatt} we show the time evolution of the first two Fourier modes (polar and nematic order) in theory and simulation, where the system was initialized at $t=0$ with nonzero polar order (all higher modes initially equal zero).
Apparently we find excellent quantitative agreement between theory and simulation.

\section{Mean field linear stability analysis of homogeneous flocking states}
Performing the mean field analysis of the letter without the assumption of spacial homogeneity, we arrive within the thermodynamic limit at the system of nonlinear Fokker-Planck equations analogous to Eq. \eqref{eq:FPmeanfield}
\begin{align}
    \partial_t P_{A/B}(\mathbf{r}, \phi)=&\Gamma_{A/B}N_{A/B}\int_0^{2\pi} d \tilde{\phi} \int_{-L/2}^{L/2}d\tilde{x} \int_{-L/2}^{L/2}d\tilde{y} \cos{\tilde{\phi}}P_{A/B}(\tilde{\mathbf{r}}, \tilde{\phi})\theta(R-|\tilde{\mathbf{r}}-\mathbf{r}|)\partial_\phi \sin(\phi)P_{A/B}(\mathbf{r}, \phi)
    \notag
    \\
    &-\Gamma_{A/B}N_{A/B}\int_0^{2\pi} d \tilde{\phi} \int_{-L/2}^{L/2}d\tilde{x} \int_{-L/2}^{L/2}d\tilde{y} \sin{\tilde{\phi}}P_{A/B}(\tilde{\mathbf{r}}, \tilde{\phi})\theta(R-|\tilde{\mathbf{r}}-\mathbf{r}|)\partial_\phi \cos(\phi)P_{A/B}(\mathbf{r}, \phi)
    \notag
    \\
    &+\Gamma_{AB}N_{B/A}\int_0^{2\pi} d \tilde{\phi} \int_{-L/2}^{L/2}d\tilde{x} \int_{-L/2}^{L/2}d\tilde{y} \cos{\tilde{\phi}}P_{B/A}(\tilde{\mathbf{r}}, \tilde{\phi})\theta(R-|\tilde{\mathbf{r}}-\mathbf{r}|)\partial_\phi \sin(\phi)P_{A/B}(\mathbf{r}, \phi)
    \notag
    \\
    &-\Gamma_{AB}N_{B/A}\int_0^{2\pi} d \tilde{\phi} \int_{-L/2}^{L/2}d\tilde{x} \int_{-L/2}^{L/2}d\tilde{y} \sin{\tilde{\phi}}P_{B/A}(\tilde{\mathbf{r}}, \tilde{\phi})\theta(R-|\tilde{\mathbf{r}}-\mathbf{r}|)\partial_\phi \cos(\phi)P_{A/B}(\mathbf{r}, \phi)
    \notag
    \\
    &-v \cos(\phi)\partial_x P_{A/B}(\mathbf{r}, \phi)-v \sin(\phi)\partial_y P_{A/B}(\mathbf{r}, \phi)+\frac{\sigma^2}{2} \partial_\phi^2 P_{A/B}(\mathbf{r}, \phi),
    \label{eq:time_evolution_full}
\end{align}
where $\theta$ denotes the Heaviside function.
Assuming a quadratic domain of area $\mathcal{A}=L\times L$ we represent the perturbation of the homogeneous stationary flocking state in Fourier space as
\begin{align}
    P_{A/B}(\mathbf{r}, \phi)=P_{A/B}^s(\phi) + \sum_{klm}\hat{f}_{klm}^{A/B} \exp\big(ik\frac{2\pi}{L}x\big)\exp\big(il\frac{2\pi}{L}y\big)\exp(i m \phi).
    \label{eq:spacial_fourier_ansatz}
\end{align}
Thus, the time evolution of the Fourier modes is given by
\begin{align}
    \partial_t \hat{f}_{klm}^{A/B}=\frac{1}{2\pi}\frac{1}{\mathcal{A}}\int_0^{2\pi}d\phi \int_{-L/2}^{L/2}d x\int_{-L/2}^{L/2}d y \bigg[\partial_t P_{A/B}(\mathbf{r}, \phi) - \partial_t P_{A/B}^s(\phi)\bigg]\exp\big(-ik\frac{2\pi}{L}x\big)\exp\big(-il\frac{2\pi}{L}y\big)\exp(-i m \phi).
    \label{eq:modes_time_evolution1}
\end{align}
Inserting the time evolution \eqref{eq:time_evolution_full} for both, $P_{A/B}$ and $P_{A/B}^s$, employing the assumed $\phi \longleftrightarrow -\phi$ symmetry of $P_{A/B}^s$ and neglecting quadratic terms in $\hat{f}$ we obtain
\begin{align}
    \partial_t \hat{f}_{klm}^{A/B}=&\frac{1}{2\pi}\frac{1}{\mathcal{A}}\int_0^{2\pi}d\phi \int_{-L/2}^{L/2}d x\int_{-L/2}^{L/2}d y \exp\big(-ik\frac{2\pi}{L}x\big)\exp\big(-il\frac{2\pi}{L}y\big)\exp(-i m \phi)
    \notag
    \\
    &\times \bigg[ \Gamma_{A/B}N_{A/B}\partial_\phi \sin\phi \sum_{qrs}\hat{f}_{qrs}^{A/B}\exp\big(iq\frac{2\pi}{L}x\big)\exp\big(ir\frac{2\pi}{L}y\big)\exp(i s \phi)
    \notag
    \\
    &\phantom{\times\bigg[++} \times\int_0^{2\pi} d \tilde{\phi} \int_{-L/2}^{L/2}d\tilde{x} \int_{-L/2}^{L/2}d\tilde{y} \cos \tilde{\phi}P_{A/B}^s(\tilde{\mathbf{r}}, \tilde{\phi})\theta(R-|\tilde{\mathbf{r}}-\mathbf{r}|)
    \notag
    \\
    &\phantom{\times\bigg[}+\Gamma_{A/B}N_{A/B}\partial_\phi \sin \phi P_{A/B}^s(\phi)
    \notag
    \\
    &\phantom{\times\bigg[++} \times\int_0^{2\pi} d \tilde{\phi} \int_{-L/2}^{L/2}d\tilde{x} \int_{-L/2}^{L/2}d\tilde{y} \cos \tilde{\phi}\sum_{qrs}\hat{f}_{qrs}^{A/B}\exp\big(iq\frac{2\pi}{L}\tilde{x}\big)\exp\big(ir\frac{2\pi}{L}\tilde{y}\big)\exp(i s \tilde{\phi})\theta(R-|\tilde{\mathbf{r}}-\mathbf{r}|)
    \notag
    \\
    &\phantom{\times\bigg[}+\Gamma_{AB}N_{B/A}\partial_{\phi}\sin\phi\sum_{qrs}\hat{f}_{qrs}^{A/B}\exp\big(iq\frac{2\pi}{L}x\big)\exp\big(ir\frac{2\pi}{L}y\big)\exp(i s \phi)
    \notag
    \\
    &\phantom{\times\bigg[++} \times\int_0^{2\pi} d \tilde{\phi} \int_{-L/2}^{L/2}d\tilde{x} \int_{-L/2}^{L/2}d\tilde{y} \cos \tilde{\phi}P_{B/A}^s(\tilde{\mathbf{r}}, \tilde{\phi})\theta(R-|\tilde{\mathbf{r}}-\mathbf{r}|)
    \notag
    \\
    &\phantom{\times\bigg[}+\Gamma_{AB}N_{B/A}\partial_\phi \sin \phi P_{A/B}^s(\phi)
    \notag
    \\
    &\phantom{\times\bigg[++} \times\int_0^{2\pi} d \tilde{\phi} \int_{-L/2}^{L/2}d\tilde{x} \int_{-L/2}^{L/2}d\tilde{y} \cos \tilde{\phi}\sum_{qrs}\hat{f}_{qrs}^{B/A}\exp\big(iq\frac{2\pi}{L}\tilde{x}\big)\exp\big(ir\frac{2\pi}{L}\tilde{y}\big)\exp(i s \tilde{\phi})\theta(R-|\tilde{\mathbf{r}}-\mathbf{r}|)
    \bigg]
    \notag
    \\
    &-\frac{iv}{2\pi}\sum_{s}\hat{f}_{kls}^{A/B}\bigg\{ k\frac{2\pi}{L}\int_0^{2\pi}d \phi \cos\phi \exp[i(s-m)\phi] +l\frac{2\pi}{L}\int_0^{2\pi}d \phi \sin\phi \exp[i(s-m)\phi]\bigg\} -\frac{\sigma^2m^2}{2}\hat{f}_{klm}^{A/B}.
    \label{eq:modes_time_evolution2}
\end{align}
Inserting the homogeneous steady state
\begin{align}
    P_{A/B}^s(\phi)=\frac{1}{\mathcal{A}} \frac{\exp(K_{A/B}\cos \phi)}{2\pi I_0(K_{A/B})}
    \label{eq:hom_steady}
\end{align}
with
\begin{align}
    K_{A/B}:=\frac{2}{\sigma^2}(\Gamma_{A/B}M_{A/B} \langle \cos \phi\rangle_{A/B}+\Gamma_{AB}M_{B/A} \langle \cos \phi\rangle_{B/A}),
\end{align}
see Eq. \eqref{eq:steadystatemf}, we arrive at
\begin{align}
    \partial_t \hat{f}_{klm}^{A/B}=&\frac{m \sigma^2}{4} K_{A/B} (\hat{f}_{k,l,m-1}^{A/B}-\hat{f}_{k,l,m+1}^{A/B})
    \notag
    \\
    &+\frac{1}{4I_0(K_{A/B})}\bigg\{ I_{1-m}(K_{A/B})+ I_{-1-m}(K_{A/B})
    +K_{A/B}\big[\frac{1}{2}I_{2-m}(K_{A/B})-I_{-m}(K_{A/B})+ \frac{1}{2}I_{-2-m}(K_{A/B})\big]
        \bigg\}
    \notag
    \\
    &\phantom{+\frac{1}{2I_0(K_{A/B})}}\times S_{kl} \bigg[ \Gamma_{A/B} M_{A/B}(\hat{f}_{k,l,-1}^{A/B}+\hat{f}_{k,l,1}^{A/B})+\Gamma_{AB} M_{B/A}(\hat{f}_{k,l,-1}^{B/A}+\hat{f}_{k,l,1}^{B/A})
    \bigg]
    \notag
    \\
    &-i\frac{vk\pi}{L}(\hat{f}_{k,l,m-1}^{A/B}+\hat{f}_{k,l,m+1}^{A/B})-\frac{vl\pi}{L}(\hat{f}_{k,l,m-1}^{A/B}-\hat{f}_{k,l,m+1}^{A/B})
    -\frac{\sigma^2m^2}{2}\hat{f}_{klm}^{A/B},
    \label{eq:modes_time_evolution3}
\end{align}
where $i$ denotes the imaginary unit, $I_\nu$ are modified Bessel functions of the first kind and
\begin{align}
    S_{kl}:=\frac{1}{\pi R^2}\int_{-L/2}^{L/2}dx\int_{-L/2}^{L/2}dy\exp\big(ik\frac{2\pi}{L}x\big)\exp\big(il\frac{2\pi}{L}y\big)\theta(R-|\mathbf{r}|)=\begin{cases}
        &1 \text{ if }k=l=0\\
        &\frac{L}{\pi R \sqrt{k^2+l^2}}J_1\big(\frac{2\pi}{L}R\sqrt{k^2+l^2}\big) \text{ else}
    \end{cases}
\end{align}
with $J_\nu$ denoting Bessel functions of the first kind.
Remarkably, different spatial modes (that means different values of $k$ and $l$) do not couple.
Thus, denoting the vector of all angular modes of A- and B- particles with spatial modes $k,l$ as $\mathbf{x}_{kl}$, we can write its time evolution as
\begin{align}
    \partial_t \mathbf{x}_{kl}= \mathcal{M}_{kl}\mathbf{x}_{kl},
    \label{eq:modes_time_evolution4}
\end{align}
where the matrix $\mathcal{M}_{kl}$ is given by Eq. \eqref{eq:modes_time_evolution3}. In Eq. \eqref{eq:modes_time_evolution4} the indexes $k$ and $l$ are fixed, that means there is no Einstein-notation used.

Within mean field theory, the stability of the homogeneous flocking state against small spatially extended perturbations depends on the sign of the real part of the eigenvalues of the matrices $\mathcal{M}_{kl}$.

\subsection{Numerical results}

The time evolution matrix $\mathcal{M}_{kl}$ is given by Eq. \eqref{eq:modes_time_evolution3}.
Given all parameters, we can solve the corresponding eigenvalue problem numerically.
Because different spatial modes do not couple the problem can be solved separately for each spatial perturbation.
Hence the corresponding matrices are not too large and the eigenvalues can be computed within a fraction of a second on a desktop computer.

We find that the homogeneous disordered solution is stable against spatial perturbations in all considered cases, where the left hand side of Eq. \eqref{eq:ptc} is positive.

Within the ordered state, where the left hand side of Eq. \eqref{eq:ptc} is negative, we always find instabilities of the homogeneous flocking phase.
Depending on parameters we find two types of instabilities.

In the first case, we find only purely longitudinal instabilities.
The corresponding orientational eigenvectors satisfy the reflection symmetry with respect to the polarization axis.
Thus, this instability does not rotate the polarization.
Parameters that exhibit the longitudinal instability are marked as blue circles in the mean field phase diagram Fig. \ref{fig:patches} $(c)$.
There are always several unstable modes.
The instability with largest growing rate is always of finite wavelength.
Thus, we find patterns (bands) of finite characteristic size.
It should be mentioned that the longitudinal mode with wavelength= system size is also slightly unstable with growing rate close to zero.
However, we suspect that those long wavelength instabilities are not relevant because nonlinear effects become  important as soon as the finite wavelength instabilities have grown significantly.

In the second case, we find only purely transversal instabilities.
The corresponding orientational eigenvectors do no satisfy the reflection symmetry with respect to the polarization axis.
Thus, the instability does rotate the polarization locally.
Parameters that exhibit this transversal instability are marked as orange squares in the mean field phase diagram Fig. \ref{fig:patches} $(c)$.
Again, the instability with largest growing rate is always of finite wavelength.

\section{Other patterns}

\begin{figure}
    \centering
    \includegraphics[width=0.23\textwidth]{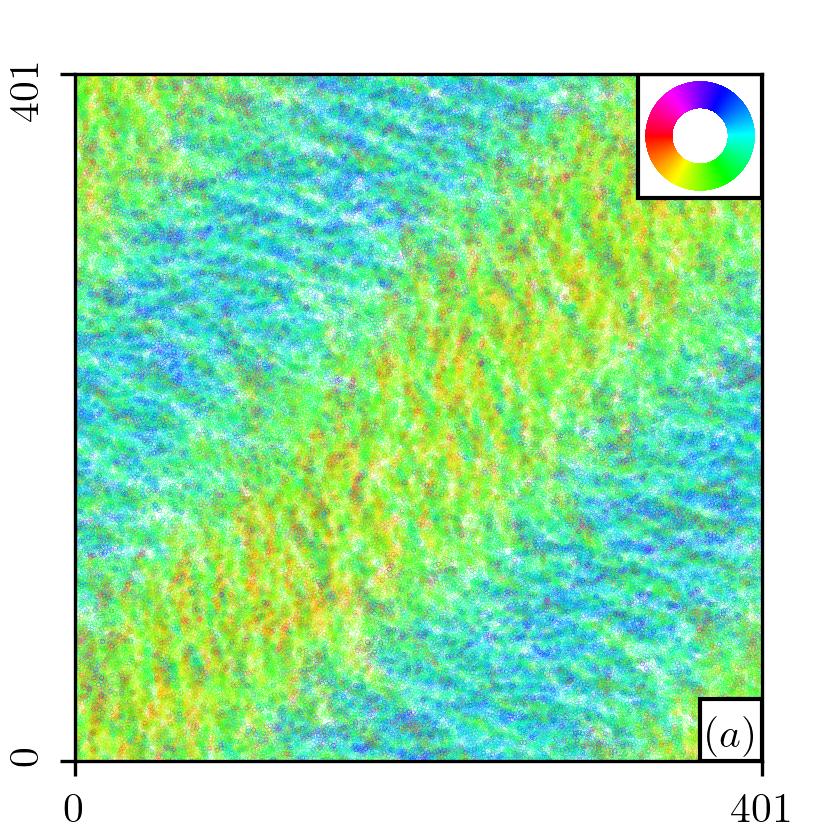}
        \includegraphics[width=0.23\textwidth]{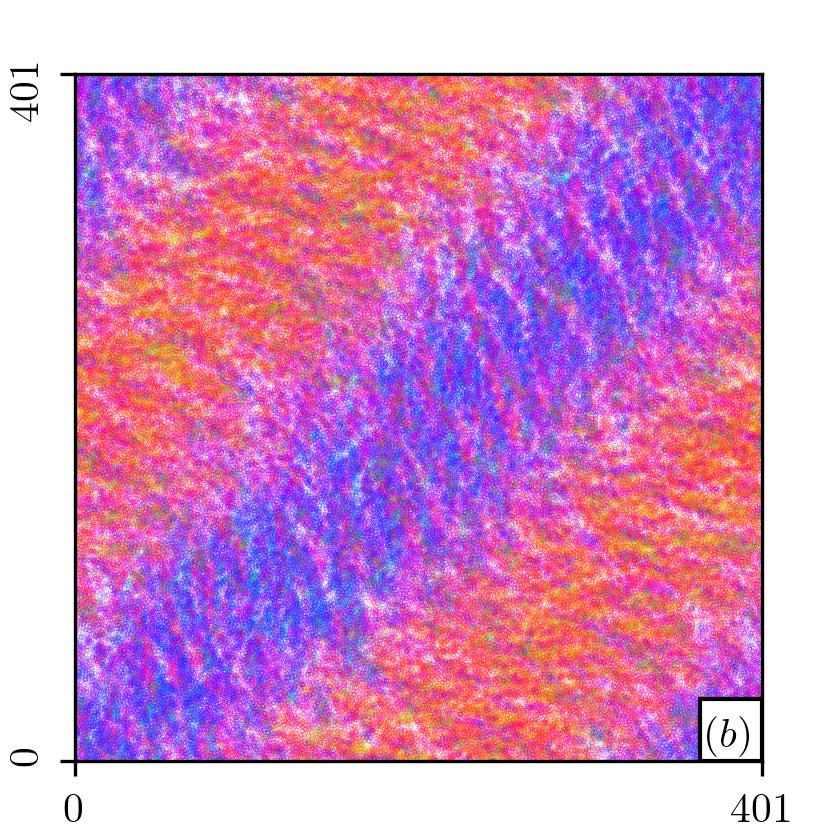}
    \caption{Snapshot of A-particles $(a)$ and B-particles $(b)$ for a different realization with the same parameters as in Fig. \ref{fig:patches} $(a),(b)$.}
    \label{fig:stripes}
\end{figure}

For the same simulation parameters as in Fig. \ref{fig:patches} $(a), (b)$ we find other, stripe-like patterns in four from ten realizations. One example is shown in Fig. \ref{fig:stripes}.
In order to distinguish which of them are the steady states we would need to either run much larger or much longer simulations, or both of it.
However, this is beyond our available computational resources.
Nevertheless we can conclude that the phase of the transversal bending instability of the homogeneous flocking state corresponds to patterns of spatially arranged regions with different local polarization. 

\begin{figure}
    \centering
    \includegraphics[width=\textwidth]{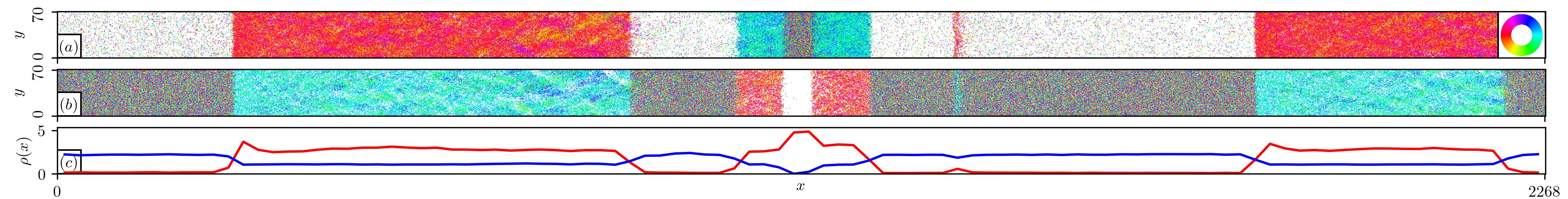}
    \caption{Snapshot of A-particles $(a)$ and B-particles $(b)$. $(c)$ Density of A-particles (red) and B-particles. The snapshot was taken from the same simulation as in Fig. \ref{fig:bands} but at an earlier time $t=5\times 10^3$. At such early times we observe long living dynamical patterns that micro-phase separate into four different phases.}
    \label{fig:dyn_pat}
\end{figure}

Within the micro-phase-separated state of travelling bands, as shown in Fig. \ref{fig:bands}, we observe long living dynamical patterns at earlier times in almost all realizations.
A snapshot of such a state is shown in Fig. \ref{fig:dyn_pat}.
We observe four different separated phases: $(ia)$ Polarized A-particles moving to the left (red) while polarized B-particles move to the right (blue). $(ib)$ Polarized A-particles moving to the right (blue) while polarized B-particles move to the left (red). $(ii)$ Disordered gas with a high density of A-particles and a very small, almost zero density of B-particles. $(iii)$ Disordered gas with a very small density of A-particles and a high density of B-particles.
In the following we shortly describe the dynamics of the interface between any pair of those phases.

\textit{Phase $(iii)$ left of phase $(ia)$.} A-particles: the polarized A-particles expand into the region of small density $(iii)$. 
B-particles: the polarized B-particles move towards the right. The high density gas $(iii)$ expands into the so created space and gets polarized towards the right.
The phase boundary moves towards the left.

\textit{Phase $(iii)$ right of phase $(ia)$.} A-particles: the polarized A-particles  move towards the left and leave almost empty disordered phase $(iii)$ behind.
B-particles: the polarized B-particles move towards the high density gas $(iii)$. At the interface they get compressed and loose polarization.
The phase boundary moves towards the left.

\textit{Phase $(ii)$ left of phase $(ia)$.} A-particles: the polarized A-particles move towards the high density region $(ii)$. At the interface they get compressed and loose polarization. 
B-particles: the polarized B-particles move towards the right leaving an almost empty disordered region behind $(ii)$.
The phase boundary moves towards the right.

\textit{Phase $(ii)$ right of phase $(ia)$.} A-particles: the polarized A-particles move away from the high density region $(ii)$. Particles from phase $(ii)$ expand into the freed space and get polarized towards the left. 
B-particles: the polarized B-particles expand into the almost empty phase $(ii)$.
The phase boundary moves towards the right.

\textit{Boundaries between phase $(ib)$ and each of $(ii)$ and $(iii)$} behave analogously to the previously discussed cases but with directions reversed.

\textit{Phase $(ia)$ left of phase $(ib)$.} A-particles: polarized particles of both phases move away from the interface and leave a low density disordered region behind.
B-particles: polarized particles of both phases collide, get compressed in the center and loose polarization.
A new phase $(iii)$ is created at the phase boundary.

\textit{Phase $(ia)$ right of phase $(ib)$.} Same as the previous case but with the roles of A- and B-particles interchanged.

\textit{Phase $(ii)$ left of phase $(iii)$.} A-particles: the high density particles from $(iii)$ expand into the low density region $(ii)$ and get polarized towards the left.
B-particles: the high density particles from $(ii)$ expand into the low density region $(iii)$ and get polarized towards the right.
At the boundary, a new phase $(ia)$ is created.

\textit{Phase $(ii)$ left of phase $(iii)$.} Analogously to the previous case but with directions reversed.

In that way the pattern changes dynamically for very long times.
When using periodic boundary conditions, as we do in simulations, the global polarization towards either the left or the right slowly increases and eventually only one of the phases $(ia), (ib)$ survives together with phase $(iii)$ resulting in a pattern shown in Fig. \ref{fig:bands}.
For the parameters of Fig. \ref{fig:bands} we find such dynamic patterns after a time $t=2.5\times 10^4$ only in two from ten realizations whereas we have the dynamic patterns in all ten realizations after short times.

\section{Simulation details}

All simulations have been performed using an Euler-Maruyama scheme implemented in the AAPPP simulation package \cite{Kuersten22}.
For Figs. \ref{fig:polar_mf}, \ref{fig:bm_theory_sim} each point is the average of ten realizations, each of them initialized uniformly and isotropic at random in a square simulation domain with periodic boundary conditions.

Fig. \ref{fig:polar_mf}: Step size is $\Delta t=5\times 10^{-3}$. The system was thermalized for $4\times 10^5$ time steps and afterwards polar order was measured for another $4\times 10^5$ time steps.

Fig. \ref{fig:bm_theory_sim} $(a)$, $v=1$: The system was thermalized for $60550000$ time steps with step size $\Delta t = 0.5$, afterwards for $30500000$ more time steps with step size $\Delta t = 0.05$ and afterwards for $30000000$ more time steps with step size $\Delta t=0.005$ and afterwards polar order was measured for another $5\times 10^5$ time steps with step size $\Delta t = 0.005$.\\
$v=3$: The system was thermalized for $152050000$ time steps with step size $\Delta t = 0.5$, afterwards for $122000000$ more time steps with step size $\Delta t = 0.05$ and afterwards for $30000000$ more time steps with step size $\Delta t=0.005$ and afterwards polar order was measured for another $5\times 10^5$ time steps with step size $\Delta t = 0.005$.\\
$v=10$, $\Gamma_A \in \{-0.0082, -0.00825, -0.0083, -0.00835, -0.0084, -0.00845, -0.0085, -0.00855, -0.0086, -0.00865,\allowbreak -0.0087, \allowbreak -0.00875, -0.0088, -0.00885, -0.0089, -0.00895, -0.00896\}$: The system was thermalized for $152050000$ time steps with step size $\Delta t = 0.5$, afterwards for $122000000$ more time steps with step size $\Delta t = 0.05$ and afterwards for $30000000$ more time steps with step size $\Delta t=0.005$ and afterwards polar order was measured for another $1\times 10^7$ time steps with step size $\Delta t = 0.005$.\\
$v=10$, $\Gamma_A \in \{-0.00897, -0.00898, -0.00899, -0.009, -0.00901, -0.00902, -0.00903, -0.00904, -0.00905, -0.00906,\allowbreak -0.00907, \allowbreak -0.0091\}$: The system was thermalized for $152050000$ time steps with step size $\Delta t = 0.5$, afterwards for $366000000$ more time steps with step size $\Delta t = 0.05$ and afterwards for $30000000$ more time steps with step size $\Delta t=0.005$ and afterwards polar order was measured for another $1\times 10^7$ time steps with step size $\Delta t = 0.005$.

Fig. \ref{fig:bm_theory_sim} $(b)$: The system was thermalized for $91050000$ time steps with step size $\Delta t = 0.5$, afterwards for $152000000$ more time steps with step size $\Delta t = 0.05$ and afterwards polar order was measured for another $5\times 10^5$ time steps with step size $\Delta t = 0.05$.

Fig. \ref{fig:bands}: Physical parameters: $M_A=M_B=5$, $N_A=N_B=256000$, $\sigma=1$, $R=1$, $v=10$, $\Gamma_A=-0.1$, $\Gamma_{AB}=\Gamma_B=-1$, $L_x=4\times \sqrt{2\pi N_A/M_A}\approx 2268$, $L_y=0.25\times\sqrt{0.5 \pi N_A/M_A}\approx 70$. Particles have been initialized uniformly and isotropically at random. The snapshot was taken after $2.5\times 10^6$ time steps of size $\Delta t = 10^{-2}$.

Fig. \ref{fig:patches} $(a-b)$: Physical parameters: $M_A=M_B=5$, $N_A=N_B=256000$, $\sigma=1$, $R=1$, $v=10$, $\Gamma_A=\Gamma_B=-1$, $\Gamma_{AB}=-1.7$, $L_x=L_y=\sqrt{\pi N_A/M_A}\approx 401$. Particles have been initialized uniformly and isotropically at random. The snapshot was taken after $1.6\times 10^6$ time steps of size $\Delta t = 10^{-2}$.
\end{appendix}

\end{document}